\documentclass[12pt]{article}
\usepackage{amsmath}
\usepackage[nosort]{cite}

\csname @addtoreset\endcsname{equation}{section}

\newcommand{\eq}{\begin{equation}}
\newcommand{\feq}{\end{equation}}
\newcommand{\eqn}{\begin{eqnarray}}
\newcommand{\feqn}{\end{eqnarray}}
\newcommand{\arr}{\begin{eqnarray*}}
\newcommand{\farr}{\end{eqnarray*}}

\newcommand{\fft}[2]{{\frac{#1}{#2}}}
\newcommand{\ft}[2]{{\textstyle\frac{#1}{#2}}}
\long
\def\omit#1{}

\DeclareMathAlphabet{\mathpzc}{OT1}{pzc}{m}{it}

\begin{document}
\baselineskip18pt

\begin{titlepage}

\begin{flushright}
MCTP-08-53
\end{flushright}

\vspace*{1cm}

\begin{center}
{\Large\bf Hamilton-Jacobi Counterterms for\\[8pt]
Einstein-Gauss-Bonnet Gravity}
\vskip 1.5cm
      James T. Liu$^{1}$ and Wafic A. Sabra$^2$\\
\vskip .6cm

      \begin{small}
      $^1$\textit{Michigan Center for Theoretical Physics,\\
        Randall Laboratory of Physics, The University of Michigan,\\
         Ann Arbor, MI 48109-1040, USA.\\}
      \end{small}
\vspace*{.6cm}

      \begin{small}
      $^2$\textit{Centre for Advanced Mathematical Sciences and
        Physics Department, \\
        American University of Beirut, Lebanon\\}
      \end{small}
\end{center}

\vfill

\begin{center}
\textbf{Abstract}
\end{center}

\begin{quote}
The on-shell gravitational action and the boundary stress tensor are
essential ingredients in the study of black hole thermodynamics. We
employ the Hamilton-Jacobi method to calculate the boundary counterterms
necessary to remove the divergences and allow the study of the
thermodynamics of Einstein-Gauss-Bonnet black holes.
\end{quote}

\vfill

\end{titlepage}

\section{Introduction}

The AdS/CFT conjecture has led to a renewed interest in the study of black hole
thermodynamics. In this new framework, thermal properties of an AdS black
hole configuration are dual to that of the finite temperature CFT. \ An
important example is the Hawking-Page phase transition \cite{Hawking:1982dh}
for black holes in AdS which corresponds to a deconfinement transition in
the dual field theory \cite{Witten:1998zw}. In order to study black hole
thermodynamics, it is standard to evaluate the on-shell gravitational
action and the boundary stress tensor.  The on-shell value of the action
(which we denote as $\Gamma$) is identified with the thermodynamic
potential $\Omega$ according to $\Gamma=\beta\Omega$. Moreover, for static
backgrounds with the time-like Killing vector $\partial/\partial t$, the
energy $E$ is
given by the ADM mass, extracted from the $tt$ component of the boundary
stress tensor. Though one expects that the thermodynamical laws are satisfied
in general, an important complication is that both $\Omega$ and $E$ are
divergent quantities and require regularization.

An approach to regularization suggested in \cite{Brown:1992br} is to
subtract the action of a reference spacetime from the action for the
spacetime of interest.  Under appropriate matching conditions, the
divergences in both actions will cancel, thus leading to finite quantities
of interest.  Although this approach is useful in many cases, it becomes
problematic when the appropriate reference background can not be found, or
when there is a potential ambiguity in the matching conditions.  In the
framework of AdS/CFT, an alternative method for removing infinities was
developed in \cite{Henningson:1998gx,Balasubramanian:1999re,Emparan:1999pm}.
Inspired by renormalization in the dual CFT, this method involves the addition
of a set of covariant boundary counterterms that remove all power law
divergences from the on-shell action.  While the values of the counterterms
were originally chosen simply to remove divergences, a subsequent refinement
of holographic renormalization came about when it was realized that the
Hamilton-Jacobi formalism may be used to determine the structure and
normalization of these counterterms \cite{deBoer:1999xf}.

Black hole thermodynamics in pure Einstein gravity with a cosmological
constant has been extensively studied, especially in light of the
AdS/CFT correspondence.  In its simplest form, this correspondence relates
$\mathcal N=4$ $SU(N)$ super-Yang-Mills theory in four dimensions with IIB
string theory on AdS$_5\times S^5$.  In the limit of large $N$ and
infinite 't~Hooft coupling, the gravitational dual simply reduces to
$\mathcal N=8$ gauged supergravity in five dimensions.  Motivated by
this AdS/CFT picture, the Hamilton-Jacobi formalism
\cite{deBoer:1999xf,Martelli:2002sp,Papadimitriou:2004ap}
was employed in \cite{hjus} to study the thermodynamics of
asymptotically AdS black holes in various dimensions, $d=4,5,6,7.$

{}From an AdS/CFT perspective, it is of natural interest to examine the
finite 't~Hooft coupling corrections to the familiar infinite coupling
results.  These corrections originate from higher derivative terms
in the $\alpha'$ expansion of the string effective action; in the
gravitational sector, they take the form $\alpha'^n R^{n+1}$ where $R$
corresponds to the Riemann tensor and its contractions.  While the
first corrections in the maximally supersymmetric ({\it i.e.}~Type II)
theories do not enter until the $\alpha'^3R^4$ order, generically the
first non-trivial terms show up at the curvature-squared level
\begin{equation}
e^{-1}\delta\mathcal L=\alpha_1R^2+\alpha_2 R_{\mu\nu}^2+\alpha_3
R_{\mu\nu\rho\sigma}^2.
\end{equation}
By making an appropriate field redefinition of the form
$g_{\mu\nu}\to g_{\mu\nu}+aR_{\mu\nu}+bg_{\mu\nu}R$, we may shift the
coefficients $\alpha_1$ and $\alpha_2$ to arbitrary values.  Thus only
$\alpha_3$ may affect physical observables.  This allows us to take
the Gauss-Bonnet combination
\begin{equation}
e^{-1}\delta\mathcal L=\alpha(R^2-4R_{\mu\nu}^2+R_{\mu\nu\rho\sigma}^2),
\end{equation}
which is the unique combination of curvature squared terms which does not
propagate ghosts \cite{Lovelock:1971yv,Zwiebach:1985uq}.  Of course, the
presence of ghosts (whose effects do not show up until the string scale)
is not a major concern at the effective supergravity level, where the
complete string theory serves as an appropriate UV completion.  Nevertheless,
the Gauss-Bonnet combination is particularly amenable to holographic
renormalization and the study of boundary field theories as it admits a
well-defined Cauchy problem for radial evolution.

It is the purpose of our present work to apply holographic renormalization
to the Einstein-Gauss-Bonnet action, and in particular to apply the
Hamilton-Jacobi method to derive a set of universal counterterms renormalizing
this action.  Local counterterms for higher derivative gravities, including
the Gauss-Bonnet combination, have previously been considered in
\cite{Nojiri:1999nd,Fukuma:2001uf,Nojiri:2001fa,Nojiri:2001ae,Cvetic:2001bk,Padilla:2003qi}.  In addition, a complementary `Kounterterm'
regularization scheme was developed in
\cite{Mora:2004kb,Olea:2005gb,Kofinas:2006hr,Olea:2006vd,Miskovic:2007mg,Kofinas:2007ns}, which involves the introduction of boundary
counterterms built out of the extrinsic curvature tensor.  (This approach
is more naturally associated with a variational principle where the
extrinsic curvature is kept fixed on the boundary.)

We organize our work as follows. In the following section, we review the
Hamiltonian formulation (for radial evolution) and we evaluate the
Hamiltonian of Einstein-Gauss-Bonnet theory to first order in $\alpha$,
the coefficient of the Gauss-Bonnet term in the bulk action. In section
three, we derive the Hamilton-Jacobi counterterms, and in section four
we compare our results to previous investigations of Gauss-Bonnet black
hole thermodynamics.  Finally, we conclude with a discussion in section
five.

\section{Einstein-Gauss-Bonnet gravity}

It is well known that, in general, higher-curvature gravitational actions
lead to potentially undesirable features such as ghosts as well as
difficulties in formulating the Cauchy problem because of the appearance of
higher-order derivatives of the metric. However, as shown by Lovelock, these
difficulties may be avoided by taking particular combinations of the
higher-curvature terms corresponding to $d$-dimensional continuations of the
lower dimensional Euler densities \cite{Lovelock:1971yv}. The family of
Lovelock actions then take the form
\begin{equation}
S_{\mathrm{bulk}}=\sum_{k=0}^{d/2}\alpha_kS_{\mathrm{bulk}}^{(k)},
\label{eq:lovelock}
\end{equation}
where \cite{Lovelock:1971yv,Teitelboim}
\begin{equation}
S_{\mathrm{bulk}}^{(k)}=-{\frac{1}{2k!}}\int_{\mathcal{M}}d^dx\sqrt{-g}
\delta^{[\mu_1\cdots\mu_{2k}]}{}_{[\nu_1\cdots\nu_{2k}]}
R^{\nu_1\nu_2}{}_{\mu_1\mu_2}\cdots
R^{\nu_{2k-1}\nu_{2k}}{}_{\mu_{2k-1}\mu_{2k}}.
\end{equation}
Note that we have included $k=0$, corresponding to a possible cosmological
constant. In particular, the first few terms in the expansion of the
Lovelock action give
\begin{equation}
S_{\mathrm{bulk}}=-\int_{\mathcal{M}}d^dx\sqrt{-g}[\alpha_0+\alpha_1 R
+\alpha_2(R^2-4R_{\mu\nu}^2+R_{\mu\nu\rho\sigma}^2)+\cdots].
\label{eq:ll3}
\end{equation}
Truncating the Lovelock theory at this level gives what may be referred to
as Einstein-Gauss-Bonnet gravity.

Since the Lovelock theory gives rise to equations of motion involving no
higher than second derivatives of the metric, it is possible to formulate a
well-defined variational principle by adding to (\ref{eq:lovelock}) a set of
generalized Gibbons-Hawking surface terms
\begin{equation}
S_{\mathrm{GH}}=\sum_{k=1}^{d/2}\alpha_kS_{\mathrm{GH}}^{(k)}.
\label{eq:ghst}
\end{equation}
In particular
\begin{equation}
S_{\mathrm{GH}}^{(1)}=-2\int_{\partial\mathcal{M}}d^{d-1}x\sqrt{-h}K
\end{equation}
is the usual Gibbons-Hawking term and
\begin{equation}
S_{\mathrm{GH}}^{(2)}=4\int_{\partial\mathcal{M}}d^{d-1}x\sqrt{-h}
[2\mathcal{G}_{ab} K^{ab}+\ft13(K^3-3KK_{ab}^2+2K_a^bK_b^cK_c^a)]
\end{equation}
is a generalized Gibbons-Hawking term \cite{Myers:1987yn,Teitelboim}. (Of
course no boundary term is needed for the $k=0$ cosmological constant term.)
Here we have assumed that the spacetime is foliated with constant $r$
hypersurfaces, orthogonal to a spacelike unit normal $n^\mu$. The boundary
metric is given by $h_{\mu\nu} =g_{\mu\nu}-n_\mu n_\nu$, and $K_{\mu\nu}$ is
the extrinsic curvature tensor defined by $K_{\mu\nu}=\nabla_{(\mu}n_{\nu)}$.
In addition, $\mathcal{G}_{ab}$ is the Einstein tensor constructed from
the boundary metric, $\mathcal{G}_{ab}=\mathcal{R}_{ab}
-\fft12h_{ab}\mathcal{R}$.

\subsection{Hamiltonian formulation}

Consistent with the foliation of spacetime with constant $r$ hypersurfaces,
we may take the above Lovelock action and derive the corresponding
Hamiltonian for radial evolution. This was in fact done in \cite{Teitelboim}
for the case of time evolution (which is easily related to radial evolution
by an appropriate analytic continuation). Although the following results are
contained in \cite{Teitelboim}, we nevertheless provide some details for
clarity of exposition.

To derive the Hamiltonian, we first use the Gauss-Codacci equations for the
$r$-foliation to rewrite the action (\ref{eq:lovelock}) in terms of
invariants built from the intrinsic and extrinsic curvatures
$\mathcal{R}^a{}_{bcd}$ and $K_{ab}$. For hypersurfaces specified by a
spacelike normal, the relevant Gauss-Codacci equation is
\begin{equation}
\overline R_{\mu\nu\rho\sigma}\equiv
h_\mu^{\mu^{\prime}}h_\nu^{\nu^{\prime}}h_\rho^{\rho^{\prime}}
h_\sigma^{\sigma^{\prime}}R_{\mu^{\prime}\nu^{\prime}\rho^{\prime}
\sigma^{\prime}}=\mathcal{R}_{\mu\nu\rho\sigma}
-K_{\mu\rho}K_{\nu\sigma}+K_{\mu\sigma}K_{\nu\rho}.
\label{eq:gc}
\end{equation}
In this case, we find
\begin{eqnarray}
S_{\mathrm{bulk}}^{(1)}+S_{\mathrm{GH}}^{(1)}&=&\int_{\mathcal{M}}d^dx
\sqrt{-g} [\mathcal{R}+K^2-K_{ab}^2],  \notag \\
S_{\mathrm{bulk}}^{(2)}+S_{\mathrm{GH}}^{(2)}&=&\int_{\mathcal{M}}d^dx
\sqrt{-g} [(\mathcal{R}+K^2-K_{ab}^2)^2-4(\mathcal{R}_{ab}
+KK_{ab}-K_{ac}K_b^c)^2 \notag \\
&&\kern5em+(\mathcal{R}_{abcd}+K_{ac}K_{bd}-K_{ad}K_{bc})^2
-\ft43K^4+8K^2K_{ab}^2  \notag \\
&&\kern5em-\ft{32}{3}KK_a^bK_b^cK_c^a-4(K_{ab}^2)^2
+8K_a^bK_b^cK_c^dK_d^a].
\end{eqnarray}
Note, in particular, that the original surface terms (\ref{eq:ghst}) are
absorbed after the Gauss-Codacci rewriting of the action.

It is now straightforward, at least in principle, to derive the conjugate
momenta $\pi^{ab}$ for radial evolution. Noting that
$K_{ab}=\fft12\mathcal{L}_n h_{ab}$ (where $\mathcal L_n$ is the Lie
derivative along the spacelike normal $n^\mu$), we may use
\begin{equation}
\pi^{ab}={\frac{1}{2\sqrt{-g}}}{\frac{\delta S}{\delta K_{ab}}},
\end{equation}
to obtain the expansion
\begin{equation}
\pi_{ab}=\sum_{k=1}^{d/2}\alpha_k\pi_{ab}^{(k)},
\label{eq:pi12}
\end{equation}
where \cite{Teitelboim}
\begin{eqnarray}
\pi_{ab}^{(1)}&=&K_{ab}-h_{ab}K,  \notag \\
\pi_{ab}^{(2)}&=&-2\bigl[h_{ab}(\mathcal{R}K -2\mathcal{R}_{cd}K^{cd})
\notag \\
&&\kern2em-\mathcal{R}K_{ab}-2\mathcal{R}_{ab}K
+4\mathcal{R}_{(a}^cK_{b)c}^{\vphantom{c}}
+2\mathcal{R}_{acbd}K^{cd}\notag \\
&&\kern2em+{\textstyle\frac{1}{3}}h_{ab}(-K^3+3KK_{cd}^2-2K_c^dK_d^eK_e^c)
\notag \\
&&\kern2em+K^2K_{ab}-2KK_a^cK_{bc}-K_{ab}K_{cd}^2+2K_a^cK_c^dK_{bd}\bigr].
\label{eq:pi1pi2}
\end{eqnarray}

The above expressions for the conjugate momenta allow us to derive the
Hamiltonian density for radial evolution
\begin{equation}
\mathcal{H}=2\sqrt{-g}\,\pi^{ab}K_{ab}-\mathcal{L}.
\end{equation}
The result is especially simple when written in terms of the projected bulk
curvature $\overline R_{\mu\nu\rho\sigma}$.
Following \cite{Teitelboim}, we find
\begin{equation}
\mathcal{H}=\sum_{k=0}^{d/2}\alpha_k\mathcal{H}^{(k)},
\end{equation}
where
\begin{eqnarray}
\mathcal{H}^{(0)}&=&\sqrt{-g},  \notag \\
\mathcal{H}^{(1)}&=&\sqrt{-g}\,\overline R,  \notag \\
\mathcal{H}^{(2)}&=&\sqrt{-g}\,\bigl(\overline R^2
-4\overline R_{\mu\nu}^2 +\overline R_{\mu\nu\rho\sigma}^2\bigr).
\end{eqnarray}
Using the Gauss-Codacci equation (\ref{eq:gc}), this is equivalent to
\begin{eqnarray}
\mathcal{H}^{(0)}&=&\sqrt{-g},  \notag \\
\mathcal{H}^{(1)}&=&\sqrt{-g}\,\bigl(\mathcal{R}-K^2+K_{ab}^2),  \notag \\
\mathcal{H}^{(2)}&=&\sqrt{-g}\,\bigl[(\mathcal{R}-K^2+K_{ab}^2)^2
-4(\mathcal{R}_{ab}-KK_{ab}+K_a^cK_{bc})^2  \notag \\
&&\kern4em+(\mathcal{R}_{abcd}-K_{ac}K_{bd}+K_{ad}K_{bc})^2\bigr].
\label{eq:h1h2}
\end{eqnarray}

Ultimately, the Hamiltonian ought to be written in terms of the canonical
variables $h_{ab}$ and $\pi^{ab}$. To accomplish this, we must invert the
relation between $\pi^{ab}$ and $K^{ab}$ given by (\ref{eq:pi12}) and
(\ref{eq:pi1pi2}). It is at this stage that the individual Lovelock terms,
parameterized by $\alpha_i$, end up mixing with each other, as the inversion
is in general a non-linear problem involving all the various $\pi_{ab}^{(k)}$
simultaneously.

In order to proceed, we restrict our attention to the Einstein-Gauss-Bonnet
theory given by the first three terms of (\ref{eq:ll3}), which we rewrite as
\begin{equation}
S_{\mathrm{bulk}}=-\int_{\mathcal{M}}d^{d}x\sqrt{-g}\left[R+(d-1)(d-2)g^2
+\alpha (R^2-4R_{\mu\nu}^2+R_{\mu\nu\rho\sigma}^2)\right]
\label{eq:gbact}
\end{equation}
(where we have set $16\pi G_d=1$).
In this case, the conjugate momentum of (\ref{eq:pi12}) may be written as
\begin{equation}
\pi_{ab}=K_{ab}-h_{ab}K+\alpha\pi_{ab}^{(2)}.
\end{equation}
A simple rearrangement gives the useful expression
\begin{equation}
K_{ab}=\pi_{ab}-{\frac{1}{(d-2)}}h_{ab}\pi-\alpha\left(\pi_{ab}^{(2)}
-{\frac{1}{d-2}}h_{ab}\pi^{(2)}\right),
\label{eq:kij0}
\end{equation}
which allows us to obtain a perturbative solution for $K_{ab}$ in terms of
$\pi_{ab}$. In particular, inserting the zeroth order expression
$K_{ab}=\pi_{ab}-h_{ab}\pi/(d-2)+\mathcal{O}(\alpha)$ into (\ref{eq:pi1pi2})
gives
\begin{eqnarray}
\pi_{ab}^{(2)}&=&2\biggl[2h_{ab}\left(\mathcal{R}_{cd}\pi^{cd}
-{\frac{1}{(d-2)}}\mathcal{R}\pi\right)+\mathcal{R}\pi_{ab}
+{\frac{4}{d-2}}\mathcal{R}_{ab}\pi
-4\mathcal{R}_{(a}^c\pi_{b)c}^{\vphantom{c}}  \notag \\
&&\quad-2\mathcal{R}_{acbd}\pi^{cd} +{\frac{2}{3}}h_{ab}
\left(\pi_c^d\pi_d^e\pi_e^c-{\frac{3}{d-2}}\pi\pi_{cd}^2
+{\frac{2}{(d-2)^2}}\pi^3\right)  \notag\\
&&\quad-2\pi_{ac}\pi^{cd}\pi_{db}+{\frac{4}{d-2}}\pi_a^c\pi_{bc}\pi
+\pi_{ab}\pi_{cd}^2-{\frac{d}{(d-2)^2}}\pi_{ab}\pi^2\biggr]
+\mathcal{O}(\alpha).
\label{eq:pi2eqn}
\end{eqnarray}

We now work out the Hamiltonian to first order in $\alpha$. Using
(\ref{eq:h1h2}) and (\ref{eq:kij0}), we write
\begin{eqnarray}
\mathcal{H}&=&\sqrt{-g}\left[\mathcal{R}+(d-1)(d-2)g^2-K^2+K_{ij}^2\right]
+\alpha\mathcal{H}^{(2)}  \notag \\
&=&\sqrt{-g}\biggl[\mathcal{R}+(d-1)(d-2)g^2+\pi_{ab}^2-{\frac{1}{d-2}}\pi^2
\notag \\
&&\kern3em+\alpha\left(-2\pi^{ab}\pi^{(2)}_{ab}+{\frac{2}{d-2}}\pi\pi^{(2)}
+{\frac{\mathcal{H}^{(2)}}{\sqrt{-g}}}\right)+\mathcal{O}(\alpha^2)\biggr].
\end{eqnarray}
The lowest order expression for $\pi_{ab}^{(2)}$ is given in
(\ref{eq:pi2eqn}), while $\mathcal{H}^{(2)}$ may be obtained from
(\ref{eq:h1h2}):
\begin{eqnarray}
\mathcal{H}^{(2)}&=&\sqrt{-g}\biggl[\left(\mathcal{R}+\pi_{ab}^2
-{\frac{1}{d-2}}\pi^2\right)^2 -4\left(\mathcal{R}_{ab}+\pi_a^c\pi_{bc}
-{\frac{1}{d-2}}\pi\pi_{ab}\right)^2  \notag \\
&&\qquad+\biggl(\mathcal{R}_{abcd}-(\pi_{ac}\pi_{bd}-\pi_{ad}\pi_{bc})
+{\frac{1}{d-2}}(h_{ac}\pi_{bd}+h_{bd}\pi_{ac}-h_{ad}\pi_{bc}
-h_{bc}\pi_{ad}) \notag \\
&&\kern4em-{\frac{1}{(d-2)^2}}\pi^2(h_{ac}h_{bd}-h_{ad}h_{bc})\biggr)^2
+\mathcal{O}(\alpha)\biggr].
\label{eq:h2eqn}
\end{eqnarray}
The resulting Hamiltonian, valid to linear order in $\alpha$, then takes the
form
\begin{eqnarray}
\mathcal{H}&=&\sqrt{-g}\biggl[\mathcal{R}+(d-1)(d-2)g^2+\pi_{ab}^2
-{\frac{1}{d-2}}\pi^2 +\alpha\biggl(\mathcal{R}^2
-4\mathcal{R}_{ij}^2+\mathcal{R}_{ijkl}^2  \notag \\
&&\kern3em-{\frac{16}{d-2}}\mathcal{R}_{ab}\pi\pi^{ab}
+{\frac{2d}{(d-2)^2}}\mathcal{R}\pi^2 -2\mathcal{R}\pi_{ab}^2
+8\mathcal{R}_{ab}\pi^{bc}\pi_c^a+4\mathcal{R}_{abcd}
\pi^{ac}\pi^{bd}  \notag \\
&&\kern3em+2\pi_a^b\pi_b^c\pi_c^d\pi_d^a-(\pi_{ab}^2)^2
-{\frac{16}{3(d-2)}}\pi\pi_a^b\pi_b^c\pi_c^a
+{\frac{2d}{(d-2)^2}}\pi^2\pi_{ab}^2-{\frac{3d-4}{3(d-2)^3}}\pi^4
\biggr)  \notag \\
&&\kern3em+\mathcal{O}(\alpha^2)\biggr].
\label{eq:hamil}
\end{eqnarray}
We will use this result in the next section when deriving the
Hamilton-Jacobi counterterms which renormalize the original action
(\ref{eq:gbact}).


\section{Hamilton-Jacobi counterterms}

The Einstein-Gauss-Bonnet action (\ref{eq:gbact}) admits solutions which are
asymptotically Anti-de Sitter, with an effective `inverse AdS radius'
$g_{\mathrm{eff}}$ given by
\begin{equation}
g^2=g_{\mathrm{eff}}^2[1-\alpha(d-3)(d-4)g_{\mathrm{eff}}^2].
\end{equation}
It is well known that the on-shell action evaluated on such a background is
divergent. In particular, assuming the metric is asymptotically given by
\begin{equation}
ds^2\sim-(1+g_{\mathrm{eff}}^2r^2)dt^2+{\frac{dr^2}{1+g_{\mathrm{eff}}^2r^2}}
+r^2d\Omega_{d-2}^2,
\label{eq:aads}
\end{equation}
the leading divergence is of a power-law form, $S\sim r^{d-1}$, with
subleading divergences falling by a factor of $1/r^2$ at each order.

The divergences of the on-shell action may be removed by holographic
renormalization
\cite{Henningson:1998gx,Balasubramanian:1999re,Emparan:1999pm}. This
involves the introduction of a counterterm action of the form
\begin{equation}
S_{\mathrm{ct}}=\int_{\partial\mathcal{M}}d^{d-1}x\sqrt{-h}
\left(W+C\mathcal{R }+D\mathcal{R}^2+E\mathcal{R}_{ab}^2
+F\mathcal{R}_{abcd}^2+\cdots\right),
\label{eq:cta}
\end{equation}
so that the renormalized action
\begin{equation}
\Gamma=S-S_{\mathrm{ct}}
\label{eq:sren}
\end{equation}
remains finite on-shell. The terms in (\ref{eq:cta}) are organized as an
expansion in powers of the inverse metric. Since examination of
(\ref{eq:aads}) indicates that $h_{ab}\sim r^2$, we see that $W$ may be
chosen to cancel the leading $r^{d-1}$ divergence, $C$ to cancel the
$r^{d-3}$ divergence, and so on.

A particularly elegant method of obtaining the coefficients in the
counterterm action is to apply the Hamilton-Jacobi equation along with
diffeomorphism invariance of the theory \cite{deBoer:1999xf}. In the
last section, we have
derived the Hamiltonian $\mathcal{H}$ for radial evolution in the
Einstein-Gauss-Bonnet theory. As this corresponds to reparameterizations
of $r$, diffeomorphism invariance constrains the Hamiltonian to vanish
\begin{equation}
\mathcal{H}[\pi ^{ab},h_{ab}]=0.
\end{equation}
To obtain the Hamilton-Jacobi equation one rewrites this Hamiltonian
constraint in terms of functional derivatives of the on-shell action. In
particular, since the on-shell action is a functional of the bulk fields
evaluated at the boundary $\partial\mathcal{M}$, the momenta can be written
as
\begin{equation}
\pi ^{ab}=\frac{1}{\sqrt{-h}}\,\frac{\delta S}{\delta h_{ab}}.
\label{eq:pidef}
\end{equation}
By replacing the momenta appearing in the Hamiltonian this functional
derivative, we obtain the Hamilton-Jacobi equation
\begin{equation}
\mathcal{H}\left[{\frac{1}{\sqrt{-h}}}\frac{\delta S}{\delta h_{ab}},
h_{ab}\right]=0.
\end{equation}

Using the Hamilton-Jacobi equation, we can obtain a set of counterterms that
will remove power-law divergences from the on-shell action. In particular,
given the renormalized action (\ref{eq:sren}), the Hamilton-Jacobi equation
takes the form \cite{deBoer:1999xf,Martelli:2002sp,McNees:2005my}
\begin{equation}
\mathcal{H}[Z^{ab}+P^{ab},h_{ab}]=0,
\label{eq:hjtot}
\end{equation}
where
\begin{equation}
Z^{ab}={\frac{1}{\sqrt{-h}}}{\frac{\delta\Gamma}{\delta h_{ab}}},\qquad
P^{ab}={\frac{1}{\sqrt{-h}}}{\frac{\delta S_{\mathrm{ct}}}{\delta h_{ab}}}.
\end{equation}
The reason this is useful is that since $Z^{ab}$ is derived from the
renormalized action, any terms in (\ref{eq:hjtot}) involving $Z^{ab}$ are
finite, or at most logarithmically divergence. Thus all power-law
divergences are fully captured by the modified Hamilton-Jacobi equation
\begin{equation}
\mathcal{H}[P^{ab},h_{ab}]=0.
\label{eq:hjp}
\end{equation}
The momentum $P^{ab}$ associated with the counterterm action (\ref{eq:cta})
may be organized in an inverse metric expansion
\begin{equation}
P^{ab}=P^{ab}_{(0)}+P^{ab}_{(1)}+P^{ab}_{(2)}+\cdots,
\end{equation}
where
\begin{eqnarray}
P^{ab}_{(0)}&=&{\textstyle\frac{1}{2}}h^{ab}W,  \notag \\
P^{ab}_{(1)}&=&-C\mathcal{G}^{ab},  \notag \\
P^{ab}_{(2)}&=&{\textstyle\frac{1}{2}}h^{ab}(D\mathcal{R}^2
+E\mathcal{R}_{cd}^2 +F\mathcal{R}_{cdef}^2)-2D\mathcal{R}\mathcal{R}^{ab}
+(2D+E+2F)\mathcal{D}^a\mathcal{D}^b\mathcal{R}  \notag \\
&&-(2D+{\textstyle\frac{1}{2}}E)h^{ab}\mathcal{D}^2\mathcal{R}
-(E+4F)\mathcal{D}^2\mathcal{R}^{ab}
-2(E+2F)\mathcal{R}^{acbd}\mathcal{R}_{cd} \notag \\
&&+4F\mathcal{R}^a_c\mathcal{R}^{bc}-2F\mathcal{R}^{acde}
\mathcal{R}^b{}_{cde}.
\label{eq:pab}
\end{eqnarray}
The resulting Hamiltonian $\mathcal{H}[P^{ab},h_{ab}]$ may likewise be
expanded in powers of the inverse metric
\begin{equation}
\mathcal{H}=\mathcal{H}_{(0)}+\mathcal{H}_{(1)}+\mathcal{H}_{(2)}+\cdots.
\end{equation}
We then demand that each term $\mathcal{H}_{(i)}$ vanishes individually. In
this fashion, we end up with a set of `descent equations'
\cite{Martelli:2002sp} for the coefficients of the divergent terms in the
counterterm action (\ref{eq:cta}).

Substituting the momenta $P^{ab}$ of (\ref{eq:pab}) into the
Einstein-Gauss-Bonnet Hamiltonian (\ref{eq:hamil}) gives
\begin{eqnarray}
\mathcal{H}_{(0)}&=&(d-1)(d-2)g^2-\frac{d-1}{4(d-2)}W^2
-\alpha{\frac{(d-1)(d-3)(d-4)}{48(d-2)^3}}W^4,  \notag \\
\mathcal{H}_{(1)}&=&\mathcal{R}\left[1-{\frac{d-3}{2(d-2)}}WC+\alpha
{\frac{(d-3)(d-4)}{2(d-2)^2}}W^2\left(1-{\frac{d-3}{6(d-2)}}WC\right)
\right], \notag \\
\mathcal{H}_{(2)}&=&\mathcal{R}^2\biggl[-{\frac{d-5}{2(d-2)}}WD
-{\frac{d-1}{4(d-2)}}C^2+\alpha\biggl(1+{\frac{(d-1)(d-4)}{(d-2)^2}}WC
\notag \\
&&\kern4em-{\frac{(d-3)(d-4)(d-5)}{12(d-2)^3}}W^3D
-{\frac{(d-1)(d-3)(d-4)}{8(d-2)^3}}W^2C^2\biggr)\biggr]  \notag \\
&&+\mathcal{R}_{ab}^2\biggl[-{\frac{d-5}{2(d-2)}}WE+C^2
+\alpha\biggl(-4 -{\frac{4(d-4)}{d-2}}WC  \notag \\
&&\kern4em-{\frac{(d-3)(d-4)(d-5)}{12(d-2)^3}}W^3E
+{\frac{(d-3)(d-4)}{2(d-2)^2}}W^2C^2\biggr)\biggr]  \notag \\
&&+\mathcal{R}_{abcd}^2\biggl[-{\frac{d-5}{2(d-2)}}WF
+\alpha\biggl(1-{\frac{(d-3)(d-4)(d-5)}{12(d-2)^3}}W^3F\biggr)\biggr]
\notag \\
&&+\mathcal{D}^2\mathcal{R}\left(2(d-2)WD+{\frac{d-1}{2}}WE+2WF\right)
\left[{\frac{1}{d-2}}+\alpha{\frac{(d-3)(d-4)}{6(d-2)^3}}W^2\right].
\notag \\
\end{eqnarray}
Starting with $\mathcal{H}_{(0)}=0$, we find
\begin{equation}
W=-2(d-2)g[1-{\textstyle\frac{1}{6}}\alpha(d-3)(d-4)g^2],
\end{equation}
which is valid to linear order in $\alpha$. This solution for $W$ may then
be inserted into the expression for $\mathcal{H}_{(1)}$. In this way, we may
solve $\mathcal{H}_{(1)}=0$ to obtain
\begin{equation}
C=-{\frac{1}{(d-3)g}}[1+{\textstyle\frac{3}{2}}\alpha(d-3)(d-4)g^2].
\end{equation}
Working out the next order terms are somewhat more involved. After solving
$\mathcal{H}_{(2)}=0$, we find
\begin{eqnarray}
D&=&\frac{d-1}{4(d-2)(d-3)^2(d-5)g^3}\left[1-\alpha g^2\left(
{\frac{4(d-2)(d-3)^2}{d-1}}
+{\frac{7(d-3)(d-4)}{2}}\right)\right],  \notag \\
E&=&-\frac{1}{(d-3)^2(d-5)g^3}\left[1-\alpha g^2\left(4(d-3)^2
+{\frac{7(d-3)(d-4)}{2}}\right)\right],  \notag \\
F&=&-{\frac{1}{(d-5)g^3}}\left(\alpha g^2\right).
\end{eqnarray}
Inserting these coefficients into the counterterm action (\ref{eq:cta})
gives
\begin{eqnarray}
S_{ct}&=&-\int_{\partial\mathcal{M}}d^{d-1}x\sqrt{-h}
\biggl[2(d-2)g (1-\ft16\alpha(d-3)(d-4)g^2)  \notag \\
&&\kern4em+\frac{1}{(d-3)g}(1+\ft32\alpha(d-3)(d-4)g^2)
\mathcal{R}  \notag \\
&&\kern4em+\frac{1}{(d-3)^2(d-5)g^3}(1
-\ft72\alpha(d-3)(d-4)g^2)
\left(\mathcal{R}_{ab}^2-\frac{d-1}{4(d-2)}\mathcal{R}^2\right)  \notag \\
&&\kern4em+{\frac{\alpha}{(d-5)g}}(\mathcal{R}^2-4\mathcal{R}_{ab}^2
+\mathcal{R}_{abcd}^2)+\cdots\biggr].
\label{eq:gbct}
\end{eqnarray}
Note that this is an expansion both in $\alpha$ (of which we have kept only
up to the linear term) and powers of the inverse metric $h^{ab}$. The
explicit counterterms given above are sufficient to cancel all power law
divergences in the Einstein-Gauss-Bonnet theory up to $d=7$.  However,
the $\mathcal O(\mathcal R^3)$ terms, which we have not computed, will
yield a finite contribution in $d=7$ which is necessary for maintaining
diffeomorphism invariance in the renormalized theory \cite{McNees:2005my}.


\section{Gauss-Bonnet black holes}

In the previous section, we have derived the counterterm action
(\ref{eq:gbct}) which may be combined with the bulk action
(\ref{eq:gbact}) and the generalized Gibbons-Hawking term
\begin{equation}
S_{\mathrm{GH}}=-2\int_{\partial\mathcal{M}}d^{d-1}x\sqrt{-h}
[K-2\alpha(2\mathcal{G}_{ab}K^{ab}
+\ft13(K^3-3KK_{ab}^2+2K_a^bK_b^cK_c^a))],
\label{eq:gengh}
\end{equation}
to obtain the total renormalized action (\ref{eq:sren})
\begin{equation}
\Gamma = S_{\mathrm{bulk}} + S_{\mathrm{GH}} - S_{\mathrm{ct}}.
\end{equation}
This action may be identified with the thermodynamic potential of the
system through $\Omega=\Gamma/\beta$ where $\beta=1/T$ is the inverse
temperature. Furthermore, we may define the boundary stress tensor
\begin{equation}
T^{ab}={\frac{2}{\sqrt{-h}}}{\frac{\delta \Gamma}{\delta h_{ab}}}.
\end{equation}
Comparing this with (\ref{eq:pidef}), we see that
\begin{equation}
T^{ab}=2\pi^{ab}-2P^{ab},
\label{eq:tab}
\end{equation}
where $\pi^{ab}$ are given by (\ref{eq:pi12}) and (\ref{eq:pi1pi2}) and
$P^{ab}$ are given by (\ref{eq:pab}). The boundary stress tensor allows
us to define the conserved momentum (and in particular the energy) of the
spacetime.

The above results allow us to investigate the thermodynamics of Gauss-Bonnet
black holes
\cite{Boulware:1985wk,Wheeler:1985nh,Wiltshire:1985us,Myers:1988ze,Cai:2001dz,Cvetic:2001bk,Cho:2002hq,Deruelle:2003ps,Torii:2005xu,Torii:2005nh}.
Before proceeding, however, we note that it is straightforward
to include a canonically normalized Maxwell field, so that the bulk action
(\ref{eq:gbact}) becomes
\begin{equation}
S_{\mathrm{bulk}}=-\int_{\mathcal{M}}d^{d}x\sqrt{-g}\left[R+(d-1)(d-2)g^2
-\ft14F_{\mu\nu}^2+\alpha
(R^2-4R_{\mu\nu}^2+R_{\mu\nu\rho\sigma}^2)\right].
\label{eq:smax}
\end{equation}
To obtain an electrically charged black hole, we take
\begin{equation}
A={\frac{Q}{(d-3)r^{d-3}}}dt\qquad\Rightarrow\qquad F=
{\frac{Q}{r^{d-2}}}dt\wedge dr,
\label{eq:fans}
\end{equation}
as well as the metric ansatz
\begin{equation}
ds^2=-fdt^2+\frac{dr^2}{f}+r^2d\Sigma_{d-2,k}^2,
\label{eq:metans}
\end{equation}
where $k$ denotes the curvature of the manifold $\Sigma_{d-2,k}$
($k=1$, $0$, $-1$). Working out the `angular' components of the Einstein
equation, we find the first-order equation
\begin{eqnarray}
&&rf^{\prime}+(d-3)(f-k)-(d-1)g^2r^2+{\frac{Q^2}{2(d-2)r^{2(d-3)}}}
\notag\\
&&\kern8em-{\frac{\alpha}{r^2}}(d-3)(d-4)(f-k)
[(d-5)(f-k)+2rf^{\prime}]=0,\qquad
\label{eq:ceom}
\end{eqnarray}
which may be solved to yield
\cite{Boulware:1985wk,Wheeler:1985nh,Wiltshire:1985us,Cai:2001dz,Cvetic:2001bk}
\begin{equation}
f=k+{\frac{r^2}{2\tilde\alpha}}\left[1\mp\sqrt{1+4\tilde\alpha
\left({\frac{\mu}{r^{d-1}}}-g^2
-{\frac{Q^2}{2(d-2)(d-3)r^{2(d-2)}}}\right)}\right],
\label{eq:fexact}
\end{equation}
where $\tilde\alpha=\alpha(d-3)(d-4)$. Here $\mu$ is a non-extremality
parameter related to the black hole mass. Note that relative simplicity of
the equation of motion (\ref{eq:ceom}) and its black hole solution is a
general feature of the Lovelock actions.

While the above black hole solution is exact in the Gauss-Bonnet parameter
$\alpha$, our derivation of the Hamilton-Jacobi counterterms was restricted
to linear order in $\alpha$. We therefore expand $f$ to first order in
$\alpha$
\begin{eqnarray}
f&=&k+g^2r^2-{\frac{\mu}{r^{d-3}}}+{\frac{Q^2}{2(d-2)(d-3)r^{2(d-3)}}}
\notag \\
&&\kern4em+{\frac{\tilde\alpha}{r^2}}\left(g^2r^2-{\frac{\mu}{r^{d-3}}}
+{\frac{Q^2}{2(d-2)(d-3)r^{2(d-3)}}}\right)^2+\cdots.
\label{eq:flin}
\end{eqnarray}
Note that we have taken the `negative' branch of (\ref{eq:fexact}), as it is
the one which has a well-behaved $\alpha\to0$ limit. In what follows, all
expressions should be understood to be taken only to linear order in
$\alpha$.

In order to parameterize the Gauss-Bonnet black hole thermodynamics, we
introduce the horizon location $r_+$, defined by $f(r_+)=0$. A simple
rearrangement of (\ref{eq:flin}) then allows us to write $\mu$ in terms of
$r_+$ as
\begin{equation}
\mu=g^2r_+^{d-1}+kr_+^{d-3}+{\frac{Q^2}{2(d-2)(d-3)r_+^{d-3}}}
+\alpha k^2(d-3)(d-4)r_+^{d-5}.
\label{eq:mueqn}
\end{equation}
This will be useful in what follows. For example, the temperature may be
obtained from $T=f^{\prime}(r_+)/4\pi$, which comes from the requirement of
avoiding a conical singularity at the horizon. Taking a derivative of
(\ref{eq:flin}) and using (\ref{eq:mueqn}) to eliminate $\mu$ gives
\begin{eqnarray}
T&=&{\frac{1}{4\pi r_+}}\biggl[(d-1)g^2r_+^2+(d-3)k
-{\frac{Q^2}{2(d-2)r_+^{2(d-3)}}}  \notag \\
&&\kern4em+\alpha k(d-3)(d-4)\biggl(-2(d-1)g^2-(d-1){\frac{k}{r_+^2}}
+{\frac{Q^2}{(d-2)r_+^{2(d-2)}}}\biggr)\biggr].\qquad
\end{eqnarray}
This matches the exact expression for the Hawking temperature
\cite{Myers:1988ze,Cai:2001dz,Cvetic:2001bk,Cho:2002hq} when expanded to
linear order in $\alpha$ (as it must, since the calculation is identical).

Turning next to the entropy, it is well known that the area expression
$S=\mathcal{A}_h/4G_d$ gets modified in higher derivative gravity. In this
case, we may instead use the Wald entropy formula
\cite{Wald:1993nt,Iyer:1994ys,Iyer:1995kg}
\begin{equation}
S=-2\pi\int_h E^{\mu\nu\rho\sigma}\epsilon_{\mu\nu}\epsilon_{\rho\sigma}
d^{d-2}x,
\end{equation}
where
\begin{equation}
E^{\mu\nu\rho\sigma} = \left.{\frac{\delta S_{\mathrm{bulk}}}{\delta
R_{\mu\nu\rho\sigma}}}\right|_{g_{\mu\nu}\mathrm{~fixed}},
\end{equation}
and where $\epsilon_{\mu\nu}$ is the binormal to the horizon $h$. Taking the
action (\ref{eq:smax}), we find
\begin{eqnarray}
E^{\mu\nu\rho\sigma}&=&-\sqrt{-g}
\bigl[\ft12(g^{\mu\rho}g^{\nu\sigma}
-g^{\mu\sigma}g^{\nu\rho})
+\alpha\bigl((g^{\mu\rho}g^{\nu\sigma} -g^{\mu\sigma}g^{\nu\rho})R
\notag \\
&&\kern6em-2(g^{\mu\rho}R^{\nu\sigma}+g^{\nu\sigma}R^{\mu\rho}
-g^{\mu\sigma}R^{\nu\rho}-g^{\nu\rho}R^{\mu\sigma})
+2R^{\mu\nu\rho\sigma}\bigr)\bigr].\qquad
\end{eqnarray}
Working out the curvature components and integrating $E^{\bar t\bar r\bar
t\bar r}$ (where the overlines indicate tangent space components) over the
horizon gives the simple entropy expression
\begin{equation}
S=4\pi\omega_{d-2,k}r_+^{d-2}\left[1+2{\frac{\alpha}{r_+^2}}
k(d-2)(d-3) \right],
\label{eq:sgb}
\end{equation}
where $\omega_{d-2,k}$ is the volume of $\Sigma_{d-2,k}$ (so that
$\omega_{d-2,k}r_+^{d-2}$ is simply the `horizon area`). Since we are
working
in units of $16\pi G_d=1$, the leading term in $S$ indeed reproduces the
standard area expression. Note that this entropy expression is universal for
spherically symmetric Gauss-Bonnet black holes in that the equations of
motion were not needed for its derivation.

The Gauss-Bonnet black hole entropy was previously computed in
\cite{Myers:1988ze} by solving the free energy expression $F=E-TS$
for the entropy (where $F$ was computed from the Euclidean action)
and in \cite{Cai:2001dz} by integrating the first law $dE=TdS$.  Both
of these computations are in agreement with the Wald entropy formula
result (\ref{eq:sgb}).  Furthermore, we may see that the linearized
expression (\ref{eq:sgb}) is in fact exact in $\alpha$.

We now work out the renormalized action for the Gauss-Bonnet black hole.
Substituting in the metric ansatz (\ref{eq:metans}) as well as the gauge
field (\ref{eq:fans}), we find that the bulk action (\ref{eq:smax}) may be
expressed as a total $r$ derivative. Integrating this from the horizon to a
cutoff $r_0$ gives
\begin{eqnarray}
S_{\mathrm{bulk}}&=&\beta\omega_{d-2,k}\Bigl[-(d-2)g^2r^{d-1}
+r^{d-3}\bigl((d-2)(f-k)+rf^{\prime}\bigr)+{\frac{Q^2}{2(d-3)r^{d-3}}}
\notag \\
&&\kern6em-\alpha(d-2)(d-3)r^{d-5}(f-k)\bigl((d-4)(f-k)
+2rf^{\prime}\bigr)\Bigr]_{r_+}^{r_0},\qquad
\end{eqnarray}
where $\beta=1/T$ is the period of the timelike circle. Using the explicit
form of $f$ given in (\ref{eq:flin}) as well as the relation
(\ref{eq:mueqn}), we obtain
\begin{eqnarray}
S_{\mathrm{bulk}}&=&\beta\omega_{d-2,k}\biggl[2g^2r_0^{d-1}(1-d(d-3)\alpha
g^2) -2\mu(1-2(d-3)\alpha g^2)+2kr_+^{d-3}  \notag \\
&&\kern3em+2\alpha k(d-3)\left(-2(d-1)g^2r_+^{d-3}-2kr_+^{d-5}
+{\frac{Q^2}{(d-2)r_+^{d-1}}}\right)\biggr].
\label{eq:gbsbulk}
\end{eqnarray}
This clearly exhibits the leading power law divergence
$I_{\mathrm{bulk}}\sim 2g^2 r_0^{d-1}$. The generalized Gibbons-Hawking
term is evaluated at the cutoff surface $r=r_0$. From (\ref{eq:gengh}),
we find
\begin{eqnarray}
\label{eq:gbsgh}
S_{\mathrm{GH}}&=&\beta\omega_{d-2,k}\Bigl[-r^{d-3}\bigl(2(d-2)f
+rf^{\prime}\bigr)  \notag \\
&&\kern6em+2\alpha(d-2)(d-3)r^{d-5}\bigl((d-4)f(\ft23f-2k)
+(f-k)rf^{\prime}\bigr)\Bigr]_{r=r_0}  \notag \\
&=&\beta\omega_{d-2,k}\Bigl[-2(d-1)g^2r_0^{d-1}(1
+\ft13(d-3)(d-8)\alpha g^2)
\notag \\
&&\kern3.7em-2(d-2)kr_0^{d-3}(1+\ft23(d-3)(d-4)\alpha g^2)
\notag \\
&&\kern3.7em-\ft83\alpha(d-2)(d-3)(d-4)k^2r_0^{d-5}
+(d-1)\mu(1+\ft43(d-3)(d-5)\alpha g^2)\Bigr].
\notag \\
\end{eqnarray}
Adding together $S_{\mathrm{bulk}}$ and $S_{\mathrm{GH}}$, we see that the
power law divergences are given by $r_0^{d-1}$, $r_0^{d-3}$ and $r_0^{d-5}$
(assuming $d$ is sufficiently large, of course). These will be canceled by
the counterterm action (\ref{eq:gbct}).

Recall that the derivation of the counterterm action
involved an expansion in powers of the inverse metric $h^{ab}$
\begin{equation}
S_{\mathrm{ct}}=S_{(0)}+S_{(1)}+S_{(2)}+\cdots,
\end{equation}
where the leading divergence of $S_{(k)}$ is of the form $r_0^{d-2k}$.
Therefore, for arbitrary dimension $d$, we would need at least the first
three counterterms to cancel the divergences of $S_{\mathrm{bulk}}
+S_{\mathrm{GH}}$. For the Gauss-Bonnet black hole, we obtain from
(\ref{eq:gbct})
\begin{eqnarray}
S_{\mathrm{ct}}&=&-\beta\omega_{d-2,k}(d-2)\Bigl[2g^2r_0^{d-1}(1
+\ft13(d-3)(d-4)\alpha g^2)  \notag \\
&&\kern6em+2kr_0^{d-3}(1+\ft23(d-3)(d-4)\alpha g^2)
+\ft83\alpha(d-3)(d-4)k^2r_0^{d-5}+  \notag \\
&&\kern6em-\mu(1+\ft43(d-3)(d-4)\alpha g^2)
+\cdots\Bigr].
\label{eq:gbsct}
\end{eqnarray}
Note that here we have taken the dimensional continuation approach of
\cite{McNees:2005my}. In particular, the dimension dependent poles in
(\ref{eq:gbct}) are canceled by zeros in the boundary curvature expressions.
This allows, for example, $S_{(2)}$ to generate a finite counterterm in $d=5$
dimensions. In the two-derivative theory, this finite contribution removes
the `Casimir energy' of global AdS$_5$ and at the same time restores full
diffeomorphism invariance of the renormalized theory \cite{McNees:2005my}.

Adding together (\ref{eq:gbsbulk}), (\ref{eq:gbsgh}) and (\ref{eq:gbsct})
finally yields the renormalized thermodynamic potential
\begin{equation}
\Omega=\omega_{d-2,k}\left[-\mu+2kr_+^{d-3}+2\alpha
k(d-3)\left(-2(d-1)g^2r_+^{d-3}-2kr_+^{d-5}
+{\frac{Q^2}{(d-2)r_+^{d-1}}}\right)\right],
\end{equation}
where $\Omega=\Gamma/\beta$, and where $\mu$ is given in (\ref{eq:mueqn}).
This expression for the thermodynamic
potential agrees (at linear order in $\alpha$) with the free energy
calculations using background subtraction to regulate the Euclidean
action \cite{Myers:1988ze,Cvetic:2001bk,Cho:2002hq} and derived through
$F=E-TS$ \cite{Myers:1988ze}.  This provides a welcome check on the
counterterm coefficients in (\ref{eq:gbct}), which involved a fair
bit of manipulation to extract from the Einstein-Gauss-Bonnet action.
We wish to stress that the inclusion of the finite counterterm was
necessary in order to obtain agreement with the previous free energy
results.

The final quantity we are interested in is the energy of the system. For the
energy, we focus on the $tt$ component of the boundary stress tensor. Using
(\ref{eq:tab}) as well as
\begin{eqnarray}
\sqrt{-h}\pi ^{\bar{t}\bar{t}} &=&-(d-2)r^{d-3}f-2\alpha
r^{d-5}(d-2)(d-3)(d-4)f(k-\ft13f),  \notag \\
\sqrt{-h}P^{\bar{t}\bar{t}} &=&\ft12
\mathcal{L}_{\mathrm{ct}}
\end{eqnarray}
(where $\mathcal{L}_{\mathrm{ct}}$ is the counterterm Lagrangian of
(\ref{eq:gbct}), and where this expression holds for the constant curvature
boundary geometry $S^{1}\times \Sigma _{d-2,k}$), we obtain a simple
expression for the energy
\begin{equation}
E=\omega _{d-2,k}(d-2)\mu .
\end{equation}
We may now see that the free energy and energy are related by the standard
expression
\begin{equation}
\Omega=E-TS-\mathcal{Q}\Phi
\end{equation}
where $\mathcal{Q}=\omega _{d-2,k}Q$ is the normalized electric charge, and
\begin{equation}
\Phi =A_{t}(r_{+})-A_{t}(\infty )={\frac{Q}{(d-3)r_{+}^{d-3}}}
\end{equation}
is the electric potential at the horizon.

\section{Summary}

The calculation of the on-shell action and boundary stress tensor is
an important aspect of the study of black hole thermodynamics. Such
quantities are generally divergent and require renormalization.  While
various approaches, including background subtraction, have been developed,
holographic renormalization using the Hamilton-Jacobi formalism is
particularly elegant and useful in the study of the
thermodynamics of black holes in asymptotically AdS spacetimes. This
approach generates the appropriate boundary counterterms needed to remove
all divergences of the on-shell action for $R$-charged AdS black
holes in various dimensions.

In this paper, we have focused on the Einstein-Gauss-Bonnet system  with a
negative cosmological constant and used the Hamiltonian-Jacobi approach to
evaluate the counterterms up to linear order in $\alpha$, the coupling of the
Gauss-Bonnet term.  In general, this linear order in $\alpha$ is all that is
physically relevant when considering the $R^2$ corrections in the expansion
of the full higher-derivative effective action.  However, it is noteworthy
that the Gauss-Bonnet form of the $R^2$ action admits exact $R$-charged
black hole solutions.  Because of this, exact expressions may be obtained
for the thermodynamic quantities calculated in the previous section.  In
particular, the temperature and thermodynamic potential take the form
\cite{Myers:1988ze,Cai:2001dz,Cvetic:2001bk,Cho:2002hq}
\begin{eqnarray}
T&=&\fft1{4\pi r_+}\fft1{r_+^2+2\tilde\alpha k}\left(
(d-1)g^2r_+^4+(d-3)kr_+^2+(d-5)\tilde\alpha k^2-\fft{Q^2}{2(d-2)r_+^{2(d-4)}}
\right),\nonumber\\
\Omega&=&-\fft{\omega_{d-2,k}r_+^{d-5}}{d-4}\left((d-2)(3g^2r_+^4+kr_+^2
-\tilde\alpha k^2)-\fft{Q^2}{2(d-3)r_+^{2(d-4)}}-8\pi r_+^3T\right).
\end{eqnarray}
It would be of interest to see if the Hamilton-Jacobi method can be extended
to capture the non-linear terms as well.  We recall, however, that the
main reason we had linearized in $\alpha$ was so we could invert the
relations (\ref{eq:pi1pi2}) for the conjugate momenta in order to derive
the Hamiltonian (\ref{eq:hamil}).  Obtaining the exact Hamiltonian through
a non-linear inversion of $\pi_{ab}\leftrightarrow K_{ab}$ looks to be a
challenge.

The main reason exact solutions of the Einstein-Gauss-Bonnet theory are
available is that, while this is a higher-derivative gravitational system,
the equations of motion arising from the Gauss-Bonnet combination do not
involve higher that two derivative of the metric.  For this reason, the
Cauchy problem for radial evolution remains well defined when conventional
Dirichlet conditions are imposed on the boundary.  Other curvature
combinations such as the Weyl tensor squared combination, which naturally
arises in the higher derivative corrections to five-dimensional
$\mathcal N=2$ supergravity \cite{Hanaki:2006pj}, do not admit a well-posed
Dirichlet problem nor an appropriate generalization of the Gibbons-Hawking
term \cite{Myers:1987yn}.  This appears to be a major obstruction to
generalizing the Hamilton-Jacobi approach to holographic renormalization
to other theories with higher-curvature terms.

Finally, we note that, while a main objective of holographic renormalization
is the removal of divergences of the on-shell AdS action, the Hamilton-Jacobi
method introduces the additional framework of diffeomorphism invariance to
the construction of the counterterm action $S_{\rm ct}$.  In practice, this
provides no
additional information for the counterterms that remove power law divergences
in the action.  However, the Hamilton-Jacobi method {\it does} naturally
determine the finite counterterms which would otherwise be free (and related
to different renormalization schemes in the dual CFT).  For theories involving
scalars, the leading counterterm determined by the Hamilton-Jacobi method
looks like an effective superpotential \cite{deBoer:1999xf}, and in this
fashion, the finite part of $S_{\rm ct}$ is necessary to maintain the
supersymmetry of the boundary theory \cite{Liu:2004it,hjus}.  We have not
included any scalars in the present analysis, although we expect the
generalization to be straightforward.

Even in the absence of scalars, we have been careful to take into account
the finite counterterm which arises (in odd dimensions $d$) through the
dimensional continuation of the $R^{(d-1)/2}$ terms in $S_{\rm ct}$
\cite{McNees:2005my}.  For spherically symmetric configurations of the
ordinary two-derivative Einstein theory, this finite counterterm removes
the `Casimir energy' of the AdS background.  Since this is simply a
constant, the physical effect of this subtraction is rather minimal (at
least from the AdS/CFT point of view).  However, this subtraction appears
to be more important in higher curvature theories, as the $\mathcal O(\alpha)$
contribution to the finite counterterm can no longer be interpreted as a
simple shift in the Casimir energy.  We thus feel it is most natural to
adhere to a diffeomorphism invariant renormalization scheme, which is
naturally accomplished through the Hamilton-Jacobi method.

\section*{Acknowledgments}

This material is based upon work supported by the National Science
Foundation under grant PHY-0703017 and by the US Department of Energy under
grant DE-FG02-95ER40899.  JTL wishes to thank A. Castro, J. Davis, K. Hanaki
and P. Szepietowski for useful conversations. The authors wish to acknowledge
the hospitality of the Khuri lab at the Rockefeller University, where part
of this work was completed.


\end{document}